\documentclass[times, review, 10pt]{elsarticle}
\usepackage{bm}
\usepackage{microtype}
\usepackage[utf8]{inputenx}
\usepackage[english]{babel}
\usepackage{curve2e}
\usepackage{csquotes}
\usepackage{helvet}
\usepackage{courier}
\usepackage{type1cm}
\usepackage{makeidx}
\usepackage{graphicx}
\usepackage{graphics}
\usepackage{multicol} 
\usepackage{amsmath}
\usepackage{amsthm}
\usepackage{booktabs}
\usepackage{float}
\usepackage[hang,flushmargin]{footmisc}
\usepackage{amssymb}
\usepackage{lineno}
\usepackage{multirow}
\usepackage{wasysym}
\usepackage{tikz}
\usepackage{caption}
\usepackage{subfigure}
\usepackage{amsfonts}
\usepackage{url}
\usepackage{fancyhdr}
\usepackage{mathtools}
\usepackage{units}
\usepackage[T1]{fontenc} 
\usepackage{verbatim}
\usepackage{listings}
\usepackage{xcolor}
\usepackage{footnote}
\usepackage{enumerate}
\usepackage{ragged2e}
\usepackage{array}
\usepackage{bm}
\usepackage{calrsfs}
\usepackage{rotating}
\usepackage{tabulary}
\usepackage{blkarray}
\usepackage{bigstrut}
\usepackage{gauss}
\usepackage[ruled,vlined, linesnumbered]{algorithm2e}
\usepackage[title]{appendix}
\usepackage{dirtytalk}
\usepackage[breaklinks=true]{hyperref}

\newcommand*{\argmin}{\operatornamewithlimits{argmin}\limits}
\newcommand*{\argmax}{\operatornamewithlimits{argmax}\limits}

\newcommand*{\cmin}{c_{\text{min}}}
\newcommand*{\defeq}{\vcentcolon=}
\newproof{pf}{\textbf{Proof}}
\newproof{df}{\textbf{Definition}}
\newtheorem{theo}{Theorem}
\newtheorem{prop}{Proposition}
\newtheorem{lemma}{Lemma}
\newcommand{\algorithmfootnote}[2][\footnotesize]{%
  \let\old@algocf@finish\@algocf@finish
  \def\@algocf@finish{\old@algocf@finish
    \leavevmode\rlap{\begin{minipage}{\linewidth}
    #1#2
    \end{minipage}}%
  }%
}

\journal{Pattern Recognition}

\begin{document}
\begin{frontmatter}

\title{A Deterministic Information Bottleneck Method for Clustering Mixed-Type Data}

\author[1]{Efthymios Costa}
\author[1]{Ioanna Papatsouma}
\author[2]{Angelos Markos}
\address[1]{Department of Mathematics, Imperial College London, London, United Kingdom}
\address[2]{School of Education, Democritus University of Thrace, Greece}

\begin{abstract}
In this paper, we present an information-theoretic method for clustering mixed-type data, that is, data consisting of both continuous and categorical variables. The proposed approach extends the Information Bottleneck principle to heterogeneous data through generalised product kernels, integrating continuous, nominal, and ordinal variables within a unified optimization framework. We address the following challenges: developing a systematic bandwidth selection strategy that equalises contributions across variable types, and proposing an adaptive hyperparameter updating scheme that ensures a valid solution into a predetermined number of potentially imbalanced clusters. Through simulations on 28,800 synthetic data sets and ten publicly available benchmarks, we demonstrate that the proposed method, named DIBmix, achieves superior performance compared to four established methods (KAMILA, K-Prototypes, FAMD with K-Means, and PAM with Gower's dissimilarity). Results show DIBmix particularly excels when clusters exhibit size imbalances, data contain low or moderate cluster overlap, and categorical and continuous variables are equally represented. The method presents a significant advantage over traditional centroid-based algorithms, establishing DIBmix as a competitive and theoretically grounded alternative for mixed-type data clustering.

\end{abstract}

\begin{keyword}
Deterministic Information Bottleneck \sep  Clustering \sep  Mixed-type Data \sep  Mutual Information
\end{keyword}

\end{frontmatter}

\section{Introduction}
\label{sec:intro}

The quest for effective data reduction approaches has led to the development of numerous algorithms designed to organise data into meaningful groups based on inherent similarities. This task, known as cluster analysis -- or simply clustering within the machine learning community -- has been the focus of extensive research across numerous scientific disciplines. At its core, clustering aims to uncover hidden structures in data, enabling insights that drive decision-making and hypothesis generation.

Examples of cluster analysis applications include the life sciences, where clustering is used to identify gene expression patterns or classify diseases \cite{zhao2005data}, the social sciences, where it helps understand behavioral segments \cite{ahlquist2012model}, and market research, where it supports customer segmentation and targeted marketing \cite{punj1983cluster}. Despite its broad utility, clustering faces significant challenges, especially when applied to modern data sets that are increasingly heterogeneous.

A key challenge in clustering arises when data sets consist of mixed-type variables, which include both continuous and categorical features. For instance, patient records in healthcare often combine numerical measurements, such as blood pressure, with categorical attributes like gender or diagnosis codes. Similarly, market research data may encompass demographic information alongside purchasing patterns. This combination of variable types is referred to as mixed-type data. Categorical variables may result from discretisation for privacy preservation or may naturally lack inherent ordering. Addressing this data heterogeneity requires clustering approaches that effectively balance the contributions of continuous and categorical variables.

The literature on clustering offers several taxonomies of algorithms, categorising them based on key features such as the type of output (e.g., hard or fuzzy clustering) or the methodological approach (e.g., model-based or centroid-based). For mixed-type data, model-based approaches have been widely explored, with the Gaussian-Multinomial mixture model \cite{mclachlanpeel} standing out as a prominent example. Non-model-based approaches, often referred to as distance-based methods, provide another robust framework for clustering heterogeneous data. Among these, the K-Prototypes algorithm \cite{huang1997clustering} is particularly notable for combining the strengths of K-Means clustering for continuous variables with the K-Modes method for categorical variables. Ji et al. \cite{ji2013improved} further enhanced this approach by incorporating attribute weights in the K-Prototypes algorithm through entropy-based feature weighting, demonstrating improved performance particularly on data sets with varying feature importance.

\textcolor{black}{Building on this foundation, Rezaei and Daneshpour \cite{rezaei2023mixed} proposed a three-fold approach combining density-based initialization, threshold-based similarity measures, and a hybrid assignment strategy. While effective for smaller, feature-rich data sets, their method requires careful threshold tuning and scales poorly to large data sets with few classes. Kar et al. \cite{kar2024edmd} introduced EDMD, which uses Boltzmann's entropy to compute dissimilarities for nominal and ordinal attributes separately, though its effectiveness depends on sufficient feature diversity. Mousavi and Sehhati \cite{mousavi2023} proposed GUDMM, employing Jensen-Shannon divergence and mutual information to capture inter-attribute dependencies, but the reliance on pairwise calculations limits scalability for high-dimensional data.}

To address such dimensionality challenges, Factor Analysis for Mixed Data (FAMD) \cite{pages_multiple_2014} has proven effective as a preprocessing step, embedding heterogeneous variables into a common latent space and thereby improving the clustering process. Another widely recognised method is the KAMILA algorithm \cite{foss2016semiparametric}, which integrates K-Means clustering with a likelihood-based criterion to balance contributions from continuous and categorical features. Benchmarking studies \cite{preud2021head, costa2023benchmarking} have demonstrated the strong performance of the aforementioned methods across various data sets. For a comprehensive review of clustering methods tailored to mixed-type data, we refer to \cite{ahmad2019survey}. 

In recent years, deep learning approaches such as Deep Embedded Clustering (DEC) \cite{xie2016unsupervised} and autoencoder-based methods \cite{min2018survey} have emerged as alternatives for clustering. However, their application to mixed-type data presents challenges. Categorical variables require one-hot encoding, which increases dimensionality and may obscure cluster structures. Additionally, numerous hyperparameters make fair comparisons difficult, and these methods often lack interpretability. Given our focus on a theoretically grounded approach that natively handles continuous, nominal, and ordinal variables, we restrict comparisons to established methods with well-understood properties.


Over the past two decades, the concept of information-based clustering \cite{slonim2005information} has emerged as an alternative approach, utilising ideas from information theory to address clustering challenges, though it remains relatively underexplored compared to more traditional methods. A cornerstone of this approach is the Information Bottleneck (IB) method \cite{tishby99information}, a versatile framework that maximises mutual information between input variables and desired outputs to capture the essence of data. Building on this foundation, various extensions of the IB algorithm have been proposed. Among these, the Deterministic Information Bottleneck (DIB) method \cite{strouse2017deterministic} stands out as a compelling variant for clustering applications, offering a deterministic assignment of data points to clusters. Hierarchical clustering can also be achieved using the Agglomerative IB (AIB) algorithm \cite{slonim1999agglomerative}, while a generalised IB clustering framework that allows controlling the level of fuzziness is considered in \cite{strouse2019information}.

Motivated by the need for a robust, theoretically grounded approach to handle mixed-type data, this paper extends the DIB framework to address data heterogeneity. We propose a flexible clustering method, tailored to effectively integrate continuous and categorical variables, and provide practical guidance for hyperparameter selection. Our key contributions are as follows: (a) adapting the DIB framework to handle mixed-type data, including continuous, nominal, and ordinal variables; (b) proposing a systematic strategy for hyperparameter selection to balance the contributions of different variable types in defining the cluster structure; and (c) evaluating the proposed method, termed DIBmix, through extensive simulations and publicly available dataset applications, benchmarking its performance against established clustering techniques.

The remainder of this paper is structured as follows: \textcolor{black}{Section \ref{Sec:IBM} provides background on the Information Bottleneck method and its deterministic variant.} Section \ref{Sec:Methodology} presents DIBmix, our proposed extension for mixed-type data, detailing its theoretical framework and algorithmic implementation. Section \ref{sec:params} outlines the selection process of hyperparameter values and Section \ref{Sec:simulations} discusses the simulations performed on artificial data to benchmark the proposed method against other established clustering techniques. In Section \ref{Sec:realdata}, we apply the DIBmix method to publicly available data sets and analyse its performance. \textcolor{black}{Section \ref{Sec:num_clusters} discusses the estimation of the number of clusters using information-theoretic quantities.} Finally, Section \ref{Sec:conclusion} summarises our findings and suggests avenues for future research.




\section{\textcolor{black}{The Information Bottleneck Method}} \label{Sec:IBM}

The Information Bottleneck (IB) method, introduced in \cite{tishby99information}, provides a framework for extracting relevant information from data by maximising mutual information between inputs and desired outputs. Its application to cluster analysis, along with a deterministic variant, was later elaborated in \cite{strouse2017deterministic} and \cite{strouse2019information}. In this section, we provide an overview of the core principles underlying the IB algorithm and its deterministic version. We introduce the necessary notation, state the key assumptions, and present the mathematical details of the method.

Given two signal sources $X$ and $Y$, the (D)IB method consists of compressing $X$ via a mapping (or encoder) to obtain a representation $T$ of the data such that the latter encapsulates all the information necessary for predicting $Y$. This process is guided by a Markov constraint, $T \leftrightarrow X \leftrightarrow Y$, which indicates that $T$ can access information about $Y$ only through $X$, and vice versa. In essence, this constraint represents a conditional independence relationship, stating that $T$ is conditionally independent of $Y$ given $X$.

In the context of cluster analysis, $T$ corresponds to the compressed representation of a dataset $\mathcal{D}$ as clusters, $Y$ represents the location of each point in the $p$-dimensional mixed-attribute space, and $X$ denotes the observation indices $i = 1, \ldots, n$. The Markov constraint implies that knowing the cluster assignment ($T$) of a point in $\mathcal{D}$ does not reveal its exact location ($Y$) unless the observation index ($X$) is also provided. This relationship is illustrated as a Directed Acyclic Graph (DAG) in Figure \ref{fig:markov_graph}.

\begin{figure}[H]
    \centering
    \begin{tikzpicture}
        \node[circle, draw] (T) at (0, 0) {T};
        \node[circle, draw] (X) at (2, 0) {X};
        \node[circle, draw] (Y) at (4, 0) {Y};
        
        \draw[->, line width=0.5mm] (T) -- (X);
        \draw[->, line width=0.5mm] (X) -- (Y);
    \end{tikzpicture}
    \caption{A Directed Acyclic Graph (DAG) representing the Markov constraint $T \leftrightarrow X \leftrightarrow Y$.}
    \label{fig:markov_graph}
\end{figure}

We define the optimal DIB clustering, \( q^*(t \mid x) \), as:

\begin{equation}\label{eq:dibclust}
    q^*(t \mid x) = \argmin\limits_{q(t \mid x)} H(T) - \beta I(Y; T)
\end{equation}

\noindent where \( H(T) \) represents the entropy of \( T \), and \( I(Y; T) \) denotes the mutual information between \( Y \) and \( T \). Expression \eqref{eq:dibclust} can be seen as seeking a compressed representation of the data, \( T \), that is most informative about the location of observations, \( Y \), while simultaneously imposing a constraint on cluster sizes through the entropy term \( H(T) \).

The entropy term acts as a regularisation component, balancing the tradeoff between data compression and relevance. This ensures thorough exploration of the space of all possible partitions. The parameter \( \beta \) controls the strength of this regularisation and is carefully tuned to prevent the excessive compression that might result in some clusters being eliminated (see subsection \ref{sec:regparam}, for more details). The final partition is selected based on the cluster assignment that maximises \( I(Y; T) \), ensuring that the cluster assignment of an observation reveals substantial information about its location in the data space, and vice versa.

\section{Proposed Methodology} \label{Sec:Methodology}

We now describe how the DIBmix algorithm is implemented. For simplicity, we begin with a univariate representation; the full multivariate formulation is provided in Algorithm \ref{algo:DIB} at the end of this section. To distinguish between different types of probability functions, we denote static probability density (or mass) functions by $p( \cdot)$ and those updated iteratively during clustering by $q(\cdot)$. 

Assume there are \( C \) clusters. The distribution of \( T \), representing the cluster assignments, is modeled as a discrete random variable with \( C \) possible outcomes. Its probability mass function is given by:

\begin{equation}\label{eq:qt}
    q(c) = \mathbb{P}(T = c) = \frac{1}{n}\sum\limits_{i=1}^n \mathbb{I}(\boldsymbol{x}_i \in c), \quad c \in \{1, \ldots, C\},
\end{equation}

\noindent where \( \mathbb{I}(\cdot) \) is the indicator function. Expression \eqref{eq:qt} calculates the proportion of observations assigned to each of the \( C \) clusters. These cluster probabilities are updated iteratively until convergence.

To estimate the joint density \( p(x, y) \), we leverage the relationship \( p(x, y) = p(y \mid x)p(x) \). Since \( X \) represents the observation index, we set \( p(x) = 1/n \), ensuring equal weight for all observations. This can be adjusted if there is a reason for certain observations to be more influential, provided \( \sum_x p(x) = 1 \). Estimating \( p(y \mid x) \) typically requires knowledge of the underlying data generation process, which is often unavailable. In such cases, we employ Kernel Density Estimation (KDE) as a practical alternative.

For data sets containing both continuous and categorical variables, we use a generalised product kernel \cite{li2003nonparametric} to estimate the joint density. Let \( p_c \), \( p_u \), and \( p_o \) denote the number of continuous, unordered categorical (nominal), and ordered categorical (ordinal) variables, respectively, such that \( p_c + p_u + p_o = p \). The joint density at a \( p \)-dimensional observation \( \boldsymbol{x}^* \) is estimated as:
\begin{equation}\label{eq:mixedkde}
\begin{split}
    \hat{f}\left( \boldsymbol{x}^* \right) = & \frac{1}{n\prod\limits_{j=1}^{p_c}s_j} \sum\limits_{i=1}^n \Bigg\{ 
    \prod\limits_{j=1}^{p_c}K_c\left(\frac{x_{i,j}-x^*_j}{s_j}\right) \\
    & \times \prod\limits_{j=p_c+1}^{p_u} K_u \left(x_{i, j} = x^*_j; \lambda_j \right) \\
    & \times \prod\limits_{j=p_c+p_u+1}^{p} K_o \left(x_{i, j} = x^*_j; \nu_j \right) 
    \Bigg\},
\end{split}
\end{equation}

\noindent where \( x_{i,j} \) is the value of variable \( j \) for observation \( i \), and \( x^*_j \) is the corresponding value for \( \boldsymbol{x}^* \). The functions \( K_c \), \( K_u \), and \( K_o \) are kernel functions for continuous, unordered categorical, and ordered categorical variables, respectively.

For continuous variables, popular kernel choices include Gaussian, Epanechnikov, biweight, and rectangular kernels \cite{silverman1998dens}. In this study, we use the Gaussian kernel, but the methodology presented can also be extended to other kernel functions. For nominal features, we use the kernel by Aitchison \& Aitken \cite{aitchison1976multivariate}, and for ordinal variables, we employ the kernel by Li \& Racine \cite{li2003nonparametric}. These kernel functions are defined as:
\begin{align}\label{eq:gaussiankernel}
    K_c\left(\frac{x-x'}{s}\right) & = \frac{1}{\sqrt{2\pi}} \exp\left\{ - \frac{\left(x-x'\right)^2}{2s^2} \right\}, \quad s > 0, \\\label{eq:aitchisonaitkenkernel}
    K_u\left(x = x' ; \lambda\right) & = \begin{cases}
      1-\lambda & \text{if } x = x' \\
      \frac{\lambda}{\ell-1} & \text{otherwise}
    \end{cases}, \quad 0 \leq \lambda \leq \frac{\ell-1}{\ell},\\ \label{eq:liracinekernel}
    K_o\left(x = x' ; \nu\right) & = \begin{cases}
      1 & \text{if } x = x' \\
      \nu^{\lvert x - x' \rvert} & \text{otherwise}
    \end{cases}, \quad 0 \leq \nu \leq 1.
\end{align}

Here, \( s \), \( \lambda \), and \( \nu \) are bandwidth or smoothing parameters, while \( \ell \) is the number of levels of the categorical variable. A bandwidth selection strategy is discussed in subsection \ref{sec:bwparams}.

It is worth noting that the generalised product kernel in Expression \eqref{eq:mixedkde} does not assume variable independence, a common misconception. Instead, the kernel functions act as weighting mechanisms, and their product ensures consistent and robust density estimation, as highlighted by Racine \cite{racine_continuous_2019}.

We now revisit the problem of estimating \( p(y \mid x) \). Using the generalised product kernel from Expression \eqref{eq:mixedkde}, one could theoretically estimate the density values over a grid of points \(\mathcal{X} \subseteq \mathbb{R}^p\) and then compute the target density for each observation \(i = 1, \ldots, n\). However, such an approach becomes computationally prohibitive when dealing with high-dimensional data or numerous variables, as it entails performing many redundant calculations. 

A more efficient alternative involves directly computing the kernel weight product for each pair of observations, resulting in a matrix \(\mathbf{P}\) that depends on the smoothing parameters, \(\boldsymbol{\theta}\). This matrix is defined as follows:
\begin{align*}
\mathbf{P}_{qr}(\boldsymbol{\theta}) &= p(\boldsymbol{x}_q \mid x = r; \boldsymbol{\theta}) \\
&= \prod\limits_{j=1}^{p_c} K_c\left(\frac{x_{q,j} - x_{r,j}}{s_j}\right) \\
&\quad \times \prod\limits_{j=p_c+1}^{p_c + p_u} K_u\left(x_{q, j} = x_{r, j}; \lambda_j\right) \\
&\quad \times \prod\limits_{j=p_c+p_u+1}^{p} K_o\left(x_{q, j} = x_{r, j}; \nu_j\right), \quad q, r = 1, \ldots, n,
\end{align*}

\noindent assuming, without loss of generality, that the first \(p_c\) variables are continuous, the next \(p_u\) are nominal, and the remaining \(p_o = p - p_c - p_u\) are ordinal. 

The resulting matrix \(\mathbf{P}\) serves as a similarity matrix, where the \((q, r)\)-th entry represents the probability that, given an observation indexed by \(r\), its location in the mixed-attribute space matches the \(p\)-dimensional vector \(\boldsymbol{x}_q\). To ensure that each column of \(\mathbf{P}\) forms a valid probability vector (i.e., sums to one), we apply a column-wise scaling, resulting in the scaled matrix $\mathbf{P'}(\boldsymbol{\theta})$. 

While this scaling step removes the symmetry of \(\mathbf{P}\), the resulting matrix \(\mathbf{P'}\) can be considered a perturbed symmetric matrix, satisfying \(\|\mathbf{P'} - \mathbf{P'}^\intercal\| \leq \epsilon\) for some small \(\epsilon > 0\), where \(\|\cdot\|\) denotes the max norm. This property opens avenues for incorporating feature selection, which we briefly discuss in Section \ref{Sec:conclusion}.


Examining Expression \eqref{eq:dibclust}, the mutual information term \(I(Y; T)\) can be expressed in terms of entropy components as follows:
\[
I(Y; T) = H(Y) - H(Y \mid T).
\]

\noindent This decomposition facilitates the computation of \(I(Y; T)\), provided we have access to the cluster-conditional density \(q(y \mid t)\). By applying the law of total probability, Bayes' theorem, and the conditional independence of \(T\) and \(Y\) given \(X\), we derive:
\[
q(y \mid t) = \frac{1}{q(t)}\sum\limits_{x} q(t \mid x)p(x, y).
\]

With the necessary distributional results established, the final step in implementing the algorithm involves determining an update rule for the clustering output \(q(t \mid x)\). Through variational calculus, the minimisation problem in Expression \eqref{eq:dibclust} is shown to be equivalent to maximising the negative loss function \(\mathcal{L}(t, x)\), defined as:
\[
\mathcal{L}(t, x) = \log q(t) - \beta D_{\text{KL}}\left( p(y \mid x) \lvert\rvert q(y \mid t) \right),
\]
where \(D_{\text{KL}}( \cdot \lvert\rvert \cdot)\) represents the Kullback-Leibler divergence. A detailed derivation of this result is provided in \cite{strouse2017deterministic}.

Bringing all these components together, the DIB clustering procedure for the multivariate case is summarised in Algorithm \ref{algo:DIB}. It is worth noting that the initial cluster assignment can be chosen randomly instead of being predefined. In such cases, the algorithm can be executed for multiple random initialisations, with the final solution selected based on the highest mutual information \(I(Y; T)\).

Additionally, the bandwidth parameters 
\(s_1, \ldots, s_{p_c}\), \(\lambda_{p_c+1}, \ldots, \lambda_{p_u}\), \\
and \(\nu_{p_c+p_u+1}, \ldots, \nu_p\), along with the tradeoff parameter \(\beta\), can be optimised based on specific criteria rather than being manually specified. 
Further details on the selection process for these parameters are provided in Section \ref{sec:params}.

\noindent\begin{minipage}{\textwidth}
\renewcommand\footnoterule{}   
\begin{algorithm}[H]
    \SetKwData{Left}{left}\SetKwData{This}{this}\SetKwData{Up}{up}
    \SetKwFunction{Union}{Union}\SetKwFunction{FindCompress}{FindCompress}
    \SetKwInOut{Input}{input}\SetKwInOut{Output}{output}
    \SetKwComment{Comment}{$\triangleright$ }{}
    \Input{Data set $\mathcal{D}$ with $n$ observations, $p$ variables of mixed-type, Regularisation parameter $\beta \geq 0$, vector of bandwidths $\boldsymbol{\theta} = (s_1, \ldots, s_{p_c}, \lambda_{p_c+1}, \ldots, \lambda_{p_c+p_u}, \nu_{p_c+p_u+1}, \ldots, \nu_p)^\intercal$, number of clusters $C$, initial cluster assignment matrix $\mathbf{Q^{(0)}_{T \mid X}} \in \mathbb{R}^{C \times n}$, maximum number of iterations $m^{\text{max}}$.}
    \BlankLine
    $\mathbf{P_X} \leftarrow 1/n \times \boldsymbol{1}_n\boldsymbol{1}_n^\intercal$
    \BlankLine
    Compute perturbed similarity matrix $\mathbf{P'} = \mathbf{P_{Y \mid X}}$ using generalised product kernels with the bandwidths $\boldsymbol{\theta} = (\boldsymbol{s}^\intercal, \boldsymbol{\lambda}^\intercal, \boldsymbol{\nu}^\intercal)^\intercal$.
    \BlankLine
    $\mathbf{P_{X,Y}} \leftarrow \mathbf{P'} \odot \mathbf{P_X}$
    \BlankLine
    $\mathbf{q^{(0)}_{T}} \leftarrow 1/n \times \mathbf{Q^{(0)}_{T \mid X}}\boldsymbol{1}_n$
    \BlankLine
    $\mathbf{Q^{(0)}_{Y \mid T}} \leftarrow 1/n \times \mathbf{P'} [\mathbf{Q^{(0)}_{T \mid X}} \boldsymbol{1}_C \oslash \mathbf{q_T^{(0)}} \boldsymbol{1}_n^\intercal]^\intercal$ 
    \BlankLine
    converged $\leftarrow$ 0 \Comment*[r]{Binary indicator for convergence}
    \BlankLine
    $m \leftarrow 1$ \Comment*[r]{Set counter}
    \While{$m \leq m^\text{max}$ \textbf{and}  $\mathrm{converged} \neq 1$}{
        $\boldsymbol{\boldsymbol{\mathcal{L}}^{(m)}_{T, X}} \leftarrow \log \left( \mathbf{q^{(m-1)}_{T}} \right) - \beta \mathbf{D_{\text{KL}}}\left( \mathbf{P'} \lvert\rvert \mathbf{Q^{(m-1)}_{Y \mid T}} \right)$
        \BlankLine
        $\mathbf{Q^{(m)}_{T \mid X}} \leftarrow \mathbb{I}\left\{ (1, \ldots, C)^\intercal = \argmax\limits_{c \in \{1, \ldots, C\}} \boldsymbol{\boldsymbol{\mathcal{L}}^{(m)}_{T, X}} \right\}$\Comment*[r]{Update step}
        \BlankLine
        \If{$\mathbf{Q^{(m)}_{T \mid X}} = \mathbf{Q^{(m-1)}_{T \mid X}}$}{
            converged $\leftarrow$ 1 \Comment*[r]{Convergence check}
            \BlankLine
            \textbf{break}
        }
        $\mathbf{q^{(m)}_{T}} \leftarrow 1/n \times \mathbf{Q^{(m)}_{T \mid X}}\boldsymbol{1}_n$
        \BlankLine
        $\mathbf{Q^{(m)}_{Y \mid T}} \leftarrow 1/n \times \mathbf{P'} [\mathbf{Q^{(m)}_{T \mid X}} \boldsymbol{1}_C \oslash \mathbf{q_T^{(m)}} \boldsymbol{1}_n^\intercal]^\intercal$
    }
   \Output{Cluster assignment $\mathbf{Q^{*}_{T \mid X}}$.}
    \caption{DIBmix}
    \footnotetext{$\odot$ and $\oslash$ denote the Hadamard product and division, respectively and $\boldsymbol{1}_d$ is the $d$-dimensional vector $(d \geq 1)$ consisting of ones, i.e. $\boldsymbol{1}_d = (1, \ldots, 1)^\intercal \in \mathbb{R}^d$.}
    \label{algo:DIB}
\end{algorithm}
\end{minipage}


\textcolor{black}{For large datasets, constructing the full similarity matrix $\mathbf{P'}$ becomes computationally prohibitive as the associated cost is quadratic in the number of observations $(\mathcal{O}(n^2))$. For such cases, we recommend utilising the Nystr\"{o}m approximation, which provides a reconstruction of $\mathbf{P'}$ using a subset of $m \ll n$ randomly selected landmark points \cite{williams2000using}. This replaces dense matrix operations with low-rank approximations by assuming the factorisation $\mathbf{P'} \approx \mathbf{C} \mathbf{W}\mathbf{C}^\intercal$, where $\mathbf{C}$ and $\mathbf{W}$ denote the $n \times m$ matrix of similarities between all observations and the landmarks and the $m \times m$ similarity matrix of the landmarks, respectively. Theoretical results indicate that the effective rank of kernel matrices grows slowly; thus, choosing $m \approx \sqrt{n}$ landmarks is sufficient to preserve the relevant spectral properties of the data, while reducing the computational complexity to $\mathcal{O}(n\sqrt{n})$ \cite{bach2013sharp}. The cluster recovery performance differs negligibly for different values of $m$, while lower values render the problem much more feasible in terms of computational cost; see \ref{appen7}.}

\section{Hyperparameter Selection}\label{sec:params}

The proposed algorithm depends on \(p+1\) hyperparameters: \(\beta\), \(\boldsymbol{s} = \left( s_1, \ldots, s_{p_c}\right)^\intercal\), \(\boldsymbol{\lambda} = \left( \lambda_{p_c+1}, \ldots, \lambda_{p_c + p_u} \right)^\intercal\), and \(\boldsymbol{\nu} = \left( \nu_{p_c+p_u+1}, \ldots, \nu_{p} \right)^\intercal\). These parameters play a crucial role in determining the quality and characteristics of the clustering output, making their careful selection essential. In this section, we propose a systematic approach for choosing appropriate hyperparameter values.

\subsection{Bandwidth Selection}\label{sec:bwparams}

We begin by examining the bandwidth parameters \(\boldsymbol{\theta} = (\boldsymbol{s}^\intercal, \boldsymbol{\lambda}^\intercal, \boldsymbol{\nu}^\intercal)\), where \(\boldsymbol{s} = \left( s_1, \ldots, s_{p_c}\right)^\intercal\), \(\boldsymbol{\lambda} = \left( \lambda_{p_c+1}, \ldots, \lambda_{p_c + p_u} \right)^\intercal\), and \(\boldsymbol{\nu} = \left( \nu_{p_c+p_u+1}, \ldots, \nu_{p} \right)^\intercal\), corresponding to continuous, unordered categorical, and ordered categorical variables, respectively. Proper tuning of these hyperparameters is essential for achieving reliable clustering results. Indeed, the choice of bandwidth values is far more critical than the selection of the kernel function itself, as noted in \cite{gavin1994choice}.

One potential approach for selecting bandwidths is to use maximum likelihood or least squares cross-validation methods \cite{li2003nonparametric}. However, such methods are somewhat naive in the context of clustering, as they are primarily designed for bandwidth selection in density estimation tasks. These techniques often aim to minimise the Mean Integrated Squared Error (MISE), a common loss function in density estimation \cite{silverman1998dens}. Yet in clustering applications, the objective is not simply to achieve minimal estimation error but to ensure that the resulting density, and consequently the perturbed similarity matrix \(\mathbf{P'}\), effectively reveals the underlying cluster structure. While recent work by \cite{ghashti2024mixed} introduced Maximum Similarity Cross Validation (MSCV), a bandwidth selection method specifically designed for clustering, its applicability is limited to purely distance-based clustering algorithms. MSCV relies on a distance metric constructed from kernel product sums across different variable types, making it unsuitable for methods like DIBmix that employ alternative clustering approaches.

We seek to determine bandwidth values that provide a balance between sufficient dispersion and informative clustering, enabling clusters to be effectively distinguished. Since the perturbed similarity matrix \(\mathbf{P'}\) is constructed by combining kernel values multiplicatively, our primary objective is to control the average ratio of these product values across observations. This ensures that extreme disparities among kernel product values are avoided, which could otherwise cause the algorithm to converge to suboptimal local minima. Notice that prior to the bandwidth search process, all continuous features are standardised to unit variance to ensure commensurability; the actual bandwidth in the original non-standardised space is given by the optimal bandwidth in the standardised space times the original standard deviation of that feature. This is done to facilitate bandwidth selection by restricting the search to values in a smaller range; for instance we search in the interval $[0.1, 10]$. Different standardisation techniques, such as range standardisation or robust standardisation are discussed in \cite{hennig2013find}.

For continuous variables, the level of disparity is quantified using the \textit{average furthest-neighbour kernel ratio} \(\chi(\boldsymbol{s})\), defined as:
\begin{equation}\label{eq:fnkernelratio}
    \chi(\boldsymbol{s}) = \frac{1}{n}\sum\limits_{i=1}^n \chi_i(\boldsymbol{s}), \quad \chi_i(\boldsymbol{s}) = \frac{1}{\min\limits_{i' \neq i} \prod\limits_{j=1}^{p_c}K_c\left(\frac{x_{i,j}-x_{i',j}}{s_j}\right)}.
\end{equation}

As the name suggests, \(\chi(\boldsymbol{s})\) measures the mean ratio of kernel product values for all pairs of least similar observations, given the continuous bandwidth vector \(\boldsymbol{s}\). For categorical variables, the maximum and minimum possible values of the kernel are known and determined by the branch values in Expressions \eqref{eq:aitchisonaitkenkernel} and \eqref{eq:liracinekernel}. Analogously, we define the \textit{ratio of disagreement} for unordered and ordered categorical variables, denoted by \(\xi(\lambda)\) and \(\xi(\nu)\), respectively:
\begin{equation}\label{eq:ratioofdisagreement}
    \xi(\lambda) = \frac{(1-\lambda)(\ell-1)}{\lambda}, \quad \xi(\nu)=\frac{1}{\nu^{(\ell-1)}}.
\end{equation}

To equalise the contribution of all categorical variables, we set \(\xi(\lambda) = \xi(\nu) = \xi\), which yields:
\[
\lambda = \frac{\ell-1}{\ell+\xi-1}, \quad \nu = \xi^{1/(1-\ell)}.
\]
Typically, we ensure \(1 < \xi \leq \xi^{\text{max}}\) to preserve the distinction between kernel values for identical and differing categorical observations. We recommend setting \(\xi^{\text{max}} = 2\) to avoid excessive disparities in kernel product values.

The priority for bandwidth selection depends on the data type. Continuous variables generally contain more information about a point's location in the mixed-attribute space, so they are prioritised unless categorical variables dominate the dataset. If the continuous bandwidth is small enough to make the equalising \(\xi\) exceed \(\xi^{\text{max}}\), categorical variables are given precedence.

For continuous variables, we determine a reasonable bandwidth value \(s\) (assuming \(\boldsymbol{s} = s(1, \ldots, 1)^\intercal\)) by seeking a high degree of local smoothing. This is quantified using the \textcolor{black}{\textit{average nearest-neighbour kernel ratio}} \(\zeta(\boldsymbol{s})\), defined similarly to Expression \eqref{eq:fnkernelratio}:
\begin{equation}\label{eq:nnkernelratio}
    \zeta(\boldsymbol{s}) = \frac{1}{n}\sum\limits_{i=1}^n \zeta_i(\boldsymbol{s}), \quad \zeta_i(\boldsymbol{s}) = \frac{1}{\max\limits_{i' \neq i} \prod\limits_{j=1}^{p_c}K_c\left(\frac{x_{i,j}-x_{i',j}}{s_j}\right)}.
\end{equation}

To avoid oversmoothing, we impose an upper bound on \(\zeta(\boldsymbol{s})\), ensuring it is at least 1.1. This value has been empirically validated across various experiments.

If categorical variables are prioritised, we propose setting:
\[
\xi = \xi^{\text{max}} - \frac{p_u + p_o}{p}.
\]
This heuristic ensures that the ratio of disagreement remains within \( (1, \xi^{\text{max}}]\), while accounting for the proportion of categorical variables. To ensure consistent contributions across variable types, we enforce either \(\chi(\boldsymbol{s}) = \xi^{p_u+p_o}\) or \(\xi^{p_c} = \chi(\boldsymbol{s})\), depending on which bandwidth is determined first. Notably, selecting the bandwidth for any one variable type immediately defines the bandwidth for the remaining types.

\textcolor{black}{In order to overcome the need for constructing the full perturbed similarity matrix $\mathbf{P'}$ for large data sets, a subsampling procedure can be used for calculating $\chi(\boldsymbol{s})$ and $\zeta(\boldsymbol{s})$. More precisely, given a data subsample $b$ of size $\tilde{n} \ll n$, the quantities $\chi_b(\boldsymbol{s})$ and $\zeta_b(\boldsymbol{s})$ are computed for $b = 1, \ldots, B$ and their median values across all subsamples are taken as our estimates for $\chi(\boldsymbol{s})$ and $\zeta(\boldsymbol{s})$. The use of the median is motivated by the fact that extreme values may be encountered for highly non-representative subsamples.}

\subsection{Regularisation Parameter Selection}\label{sec:regparam}

The regularisation parameter \(\beta \geq 0\) balances relevance and compression, as shown in Expression \eqref{eq:dibclust}. Higher \(\beta\) values emphasise relevance, while lower values promote compression, with $\beta \rightarrow \infty$ corresponding to the Agglomerative IB algorithm for hierarchical clustering. In the original implementation of DIB for clustering \cite{strouse2017deterministic}, the optimal \(\beta\) was determined by plotting \(I(Y; T)\) against \(H(T)\) for various \(\beta\) values and selecting the value that corresponds to the point of maximum curvature on the curve \(I(Y; T) = f(H(T))\).

Our proposed approach diverges significantly from that of \cite{strouse2017deterministic, strouse2019information}. Notably, while the original DIB algorithm used \(\beta\) as a model selection parameter to determine both the regularisation strength and the number of clusters, our method fixes the number of clusters to \(C\). However, the algorithm may still result in empty cluster, as it seeks to minimise \(H(T)\). 

To address this issue, we propose an adaptive approach for tuning \(\beta\), where \(\beta\) is updated at each iteration instead of being held constant. This dynamic adjustment ensures the smallest cluster is retained while maintaining a partition into \(C\) clusters, effectively applying just enough regularisation at each step, so as to ensure that no cluster is being dropped. The entropy term \(H(T)\) is thus used primarily to allow for imbalanced cluster sizes, as \(H(T)\) is maximised when cluster masses are equal (see Theorem \ref{thm:max_entropy} in \ref{appen1}).

At iteration \(m\), the cluster assignment is based on maximising the negative loss function:
\begin{equation*}
    \mathcal{L}^{(m)}(t, x) = \log q^{(m-1)}(t) - \beta^{(m)} D_{\text{KL}}\left( p(y \mid x) \| q^{(m-1)}(y \mid t) \right),
\end{equation*}
where \(\beta^{(m)}\) is now iteration-specific. Transitioning from \(C\) to \(C-1\) clusters occurs when \(\beta^{(m)}\) is too small, causing the smallest cluster to vanish. To prevent this, \(\beta^{(m)}\) must be large enough to retain at least one observation in the smallest cluster, indexed by:
\begin{equation*}
    \cmin^{(m-1)} \defeq \argmin_{t \in \{1, \ldots, C\}} q^{(m-1)}(t).
\end{equation*}

To ensure this, at least one observation \(x\) must satisfy:
\begin{equation*}
    q^{(m-1)}\left(\cmin^{(m-1)} \mid x\right) = 1 \quad \text{and} \quad \mathcal{L}^{(m)}\left(\cmin^{(m-1)}, x\right) > \mathcal{L}^{(m)}(t, x), \, \forall t \neq \cmin^{(m-1)}.
\end{equation*}

Substituting the loss function and solving for \(\beta^{(m)}\), we derive the following inequality:
\begin{equation}\label{eq:betaupdate}
    \beta^{(m)} > \frac{\log q^{(m-1)}\left(\cmin^{(m-1)}\right) - \log q^{(m-1)}(t)}{D_{\text{KL}}\left( p(y \mid x) \| q^{(m-1)}(y \mid \cmin^{(m-1)}) \right) - D_{\text{KL}}\left( p(y \mid x) \| q^{(m-1)}(y \mid t) \right)},
\end{equation}
for all \(t \neq \cmin^{(m-1)}\) and for at least one \(x\) in \(\cmin^{(m-1)}\). The updated \(\beta^{(m)}\) is set to the maximum of Expression \eqref{eq:betaupdate} over all such \(x\) and \(t\), with a small offset (\(10^{-5}\)) added to ensure the inequality holds strictly.

While this approach ensures the smallest cluster is retained, it cannot guarantee non-zero masses for other clusters. If a cluster \(c \neq \cmin\) exists such that:
\begin{equation*}
    \beta^{(m)} > \max_{x : q^{(m-1)}(c \mid x) = 1} \left\{ \frac{\log q^{(m-1)}(c) - \log q^{(m-1)}(t)}{D_{\text{KL}}\left( p(y \mid x) \| q^{(m-1)}(y \mid c) \right) - D_{\text{KL}}\left( p(y \mid x) \| q^{(m-1)}(y \mid t) \right)} \right\},
\end{equation*}
for all \(t \neq c, \cmin\), then cluster \(c\) is dropped. This typically happens when \(c\) contains observations with low pairwise similarities and strong affinities to other clusters. Such issues often arise from poorly initialised cluster assignments, allowing problematic initialisations to be identified and discarded early in the process.

\section{Simulations on Artificial Data} \label{Sec:simulations}
We conducted a simulation study to evaluate the performance of the proposed DIBmix method in comparison to four leading approaches for clustering mixed-type data, identified based on prior benchmarking studies \cite{ahmad2019survey,costa2023benchmarking}. The selected methods are: KAMILA (KAy-means for MIxed LArge data) \cite{foss2016semiparametric}, K-Prototypes \cite{huang1997clustering}, Factor Analysis for Mixed Data followed by K-Means (FAMD) \cite{pages_multiple_2014}, and Partitioning Around Medoids (PAM) with Gower's dissimilarity \cite{kaufman2009finding}. Notably, K-Prototypes and PAM are centroid-based clustering techniques, FAMD incorporates a dimensionality reduction step before clustering, and KAMILA is a semi-parametric method. Since the study's objective is to evaluate clustering methods that can flexibly handle mixed-type data with minimal assumptions about the data-generating process, model-based clustering algorithms were intentionally excluded from the comparison.

The DIBmix method is implemented in the R package \texttt{IBclust}, which is available on CRAN \cite{markos_ibclust_2025}. For the comparison methods, we used the following R packages: \texttt{kamila} \cite{foss2018kamila} for implementing the KAMILA method, \texttt{clustMixType} \cite{szepannek2018clustmixtype} for K-Prototypes, \texttt{FactoMineR} \cite{le2008factominer} for the FAMD with K-Means approach, and \texttt{cluster} \cite{kaufman2009finding} for PAM with Gower's dissimilarity. The code to reproduce the simulation results can be found in \url{https://github.com/EfthymiosCosta/IBclust_Simulations}.

Following best practices for conducting benchmarking studies in cluster analysis \cite{van2023white}, we designed a full factorial experiment to systematically compare the five methods. Artificial data sets were generated with varying number of clusters (3, 5 and 8), sample size (100 and 600), number of variables (8 and 16), proportion of categorical to continuous variables (0.2, 0.5 and 0.8), overlap between clusters (0.01, 0.05, 0.10 and 0.15, corresponding to very low, low, moderate and high overlap), cluster shapes (spherical and elliptical) and cluster sizes (equal and imbalanced with one cluster including 10\% of the observations and remaining observations being equally divided across the remaining clusters).

Data generation was performed using the \texttt{MixSim} function from the \textcolor{black}{homonymous} package \cite{melnykov2012mixsim}, with each scenario replicated 50 times. For continuous data, cluster overlap is defined as in \cite{maitra2010simulating}. To incorporate nominal features, quartile-based discretisation was applied. In total, 28,800 data sets were generated. Cluster recovery was assessed using the Adjusted Rand Index (ARI) \cite{hubert1985} and the Adjusted Mutual Information (AMI) \cite{vinh_information_2010}. We only report the ARI values here for brevity; the same conclusions are drawn by considering the AMI values, which are available in the GitHub link provided earlier. 

\begin{figure}[t]
\centering
	\includegraphics[scale=0.6]{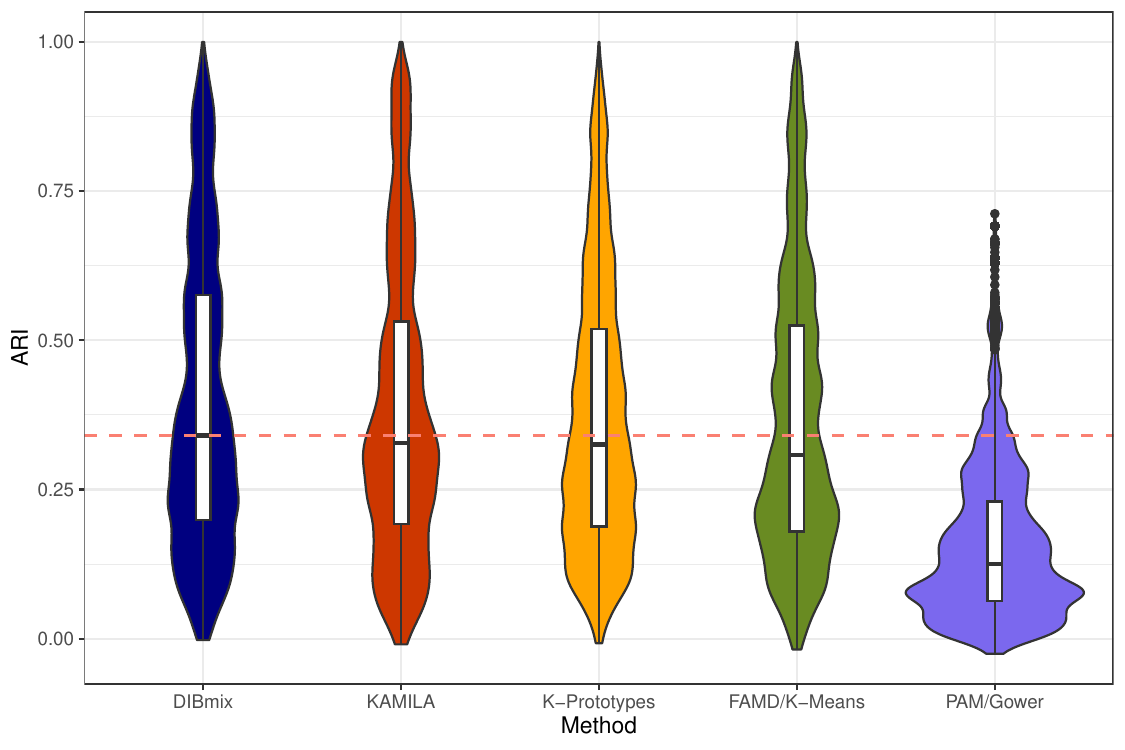}
	\caption{Violin/box plots of \textcolor{black}{ARI} values by method. \textcolor{black}{The dashed horizontal line indicates the median ARI of DIBmix ($\approx$0.340).}}
	\label{fig:violinboxplots} 
\end{figure}

The DIBmix method was implemented using the kernel functions defined in Expressions \eqref{eq:gaussiankernel}, \eqref{eq:aitchisonaitkenkernel}, and \eqref{eq:liracinekernel}. Parameter values were selected according to the strategy described in Section \ref{sec:params}. All clustering methods were initialised with 100 random starts, with a maximum of 100 iterations allowed for convergence. 

\begin{table}[!ht]
\centering
\caption{Partial $\eta^2$ values for pairwise comparisons of clustering performance (ARI, AMI) between DIBmix and competing methods.}
\label{tab:partialetasanova}
\begin{tabular}{lcccc}
\toprule
\hline
 & KAMILA & K-Prototypes & FAMD & PAM/Gower\\
\hline
\midrule
Partial $\eta^2$ (ARI) & 0.3254 & 0.6883 & 0.7788 & 0.9940 \\
Partial $\eta^2$ (AMI) & 0.4906 & 0.8295 & 0.8894 & 0.9961 \\
\hline
\bottomrule
\end{tabular}
\end{table}

Figure \ref{fig:violinboxplots} illustrates the distribution of ARI values across the five methods. The median ARI is highest for DIBmix \textcolor{black}{(0.340)}, followed by KAMILA \textcolor{black}{(0.328)}, K-Prototypes \textcolor{black}{(0.326)}, and FAMD with K-Means \textcolor{black}{(0.308)}, suggesting that these methods are more consistent in producing high-quality cluster partitions. Conversely, PAM with Gower's dissimilarity exhibits a wider spread of ARI values, including several low outliers, indicating less stable performance. A repeated measures analysis of variance in which the four competing algorithms (KAMILA, K-Prototypes, FAMD with K-Means, and PAM/Gower) are compared individually with DIBmix also verifies the improvement in cluster recovery performance with the latter. \textcolor{black}{The partial $\eta^2$ values are summarised in Table \ref{tab:partialetasanova}. These effect sizes range from 0.14 to 0.26, all exceeding or substantially exceeding Cohen's rule-of-thumb threshold of 0.14 for large effects \cite{cohen1977}. This indicates that DIBmix consistently achieves meaningfully superior cluster recovery performance compared to each competing method across the diverse simulation scenarios. Notice that $p$-values are not reported as these are almost zero due to the large number of comparisons.}

\begin{figure}[h!]
\centering
	\includegraphics[scale=0.6]{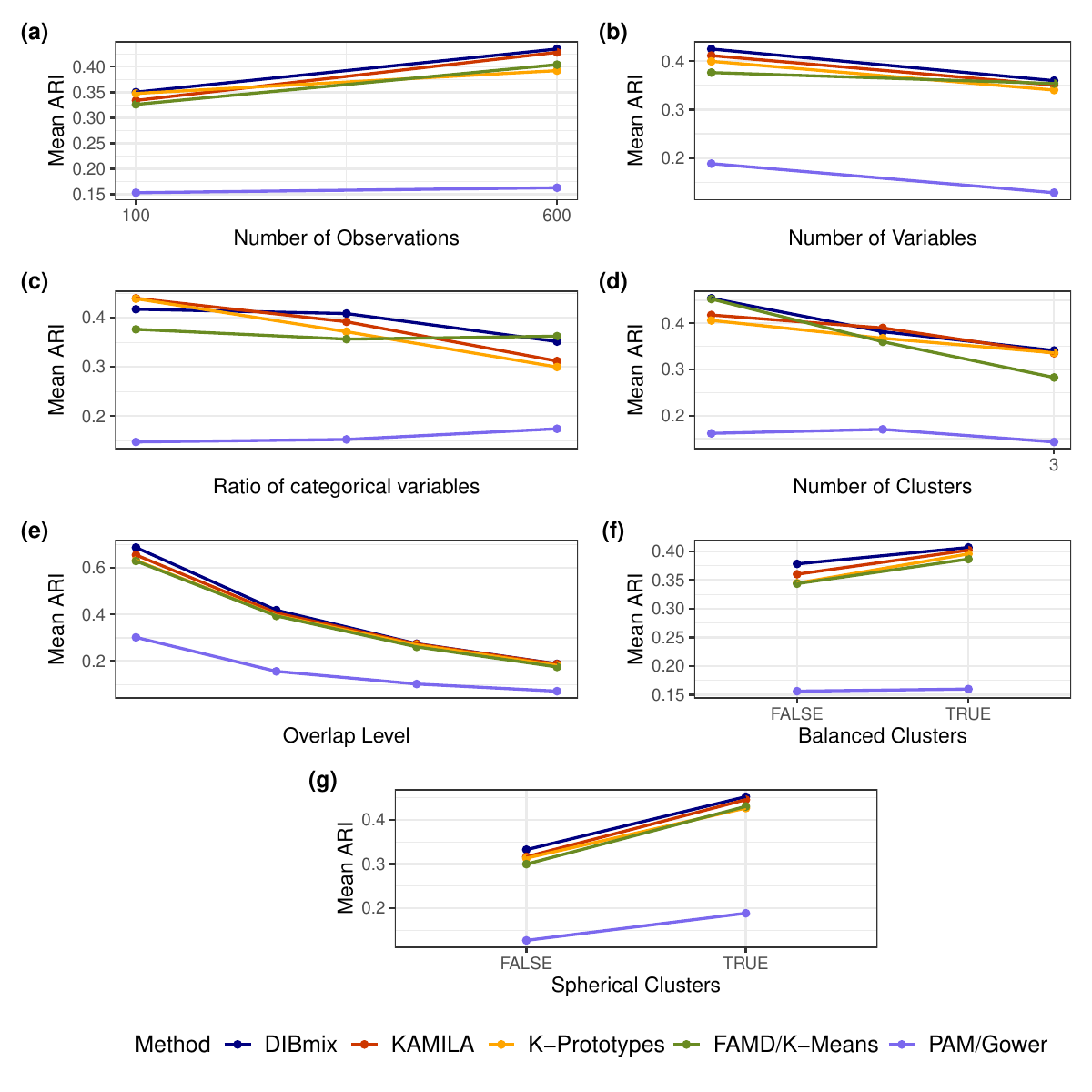}
	\caption{Mean cluster recovery in terms of ARI of the five methods under comparison across different experimental conditions}
	\label{fig:interactions} 
\end{figure}

Figure \ref{fig:interactions} provides a detailed comparison of mean ARI values across varying experimental conditions. Overall, DIBmix outperforms the other methods in a majority of scenarios. However, its performance is relatively less competitive when the proportion of categorical variables is high \textcolor{black}{(see Figure \ref{fig:interactions}(c))}. This may be attributable to the bandwidth selection strategy outlined in subsection \ref{sec:bwparams}; alternative bandwidth choices could potentially enhance performance in these cases. \textcolor{black}{Despite the concerns related to the bandwidth selection for data sets with a large proportion of categorical features}, DIBmix still performs comparably to other methods in high-categorical scenarios, where most methods, except FAMD\textcolor{black}{/K-Means}, struggle. The dimensionality reduction step in FAMD likely provides an advantage in handling data sets dominated by categorical features. Another noteworthy observation is DIBmix's robustness to cluster imbalances\textcolor{black}{, as illustrated in Figure \ref{fig:interactions}(f)}. Unlike KAMILA, K-Prototypes, and FAMD\textcolor{black}{/K-Means}, DIBmix effectively handles data sets with unbalanced cluster sizes \textcolor{black}{by allowing partitions to be highly imbalanced and not necessarily favouring equally-sized groups}. This robustness arises from the entropy term in its objective function, which minimises for imbalanced clusters, thus allowing exploration of diverse partition structures.

\section{Applications to Publicly Available Data} \label{Sec:realdata}

While artificial data provides controlled settings, it does not cover the full range of scenarios encountered in practice. We therefore complement those experiments with evaluations on publicly available data sets from the UCI Machine Learning Repository \cite{UCIRepository}. These data sets were originally designed for classification tasks, providing a benchmark for assessing clustering accuracy. The ARI values, averaged over one hundred random initialisations, are summarised in Table \ref{tab:comparison}, with the best result for each dataset shown in bold. \textcolor{black}{Note that Adult/Census Income was the only data set consisting of over a thousand observations, so a Nystr\"{o}m approximation with $m = \lceil \sqrt{n} \rceil = 174$ random landmark points and a subsampling procedure with a hundred subsamples of a thousand observations for the choice of hyperparameter values was used. Minimal differences in clustering performance were found for different values of $m$, while the recommended value of $\lceil \sqrt{n} \rceil$ required a reasonable runtime; see \ref{appen7}.}

\begin{table}[H]
\caption{Performance of five clustering methods on ten mixed-type data sets from the UCI repository (values are ARIs). Largest values appear in bold.}
\label{tab:comparison}
\scriptsize{
\begin{tabular}{lrrrrr}
\toprule
\hline
Dataset & DIBmix & KAMILA & K-Prototypes & FAMD & Gower/PAM \\
\hline
\midrule
Dermatology & 0.2698 & 0.4278 & 0.5500 & \textbf{0.7258} & 0.6130\\
($C=6$, $n = 358$, $p_c = 1$, $p_u=33$, $p_o = 0$) & & & & & \\
Heart Disease & \textbf{0.4381} & 0.3611 & 0.3775 & 0.3775 & 0.3944\\
($C=2$, $n = 297$, $p_c =6$, $p_u = 7$, $p_o = 0$) & & & & & \\
Adult/Census Income & \textbf{0.1749} & 0.1514 & 0.0864 & 0.1558 & 0.1007 \\
($C=2$, $n=30162$, $p_c = 6$, $p_u = 7$, $p_o = 1$) & & & & & \\
Hepatitis & 0.1782 & 0.2279 & 0.1101 & 0.2279 & \textbf{0.2625} \\
($C=2$, $n = 80$ $p_c =6$, $p_u = 13$, $p_o = 0$) & & & & & \\
Australian Institute of Sport & \textbf{0.8471} & \textbf{0.8471} & \textbf{0.8471} & 0.7930 & 0.3741\\
($C=2$, $n = 202$, $p_c =11$, $p_u = 1$, $p_o = 0$) & & & & & \\
Inflammation & 0.4856 & 0.4856 & \textbf{0.7486} & 0.4147 & 0.0621\\
($C=2$, $n = 120$, $p_c =1$, $p_u = 5$, $p_o = 0$) & & & & & \\
Statlog (Australian Credit Approval) & \textbf{0.4245} & 0.3009 & 0.3872 & 0.3761 & 0.4092\\
($C=2$, $n=690$, $p_c =6$, $p_u = 8$, $p_o = 0$) & & & & & \\
Credit Approval & 0.3357 & 0.2868 & \textbf{0.3685} & 0.0013 & 0.3430\\
($C=2$, $n=653$, $p_c =6$, $p_u = 9$, $p_o = 0$) & & & & & \\
Echocardiogram & 0.3352 & 0.1840 & \textbf{0.3815} & 0.2700 & 0.0396\\
($C=2$, $n = 61$, $p_c = 7$, $p_u = 2$, $p_o = 0$) & & & & & \\
Byar Prostate Cancer & 0.2558 & \textbf{0.3941} & 0.1278 & 0.2984 & 0.0363\\
($C=2$, $n = 475$, $p_c = 8$, $p_u = 4$, $p_o = 1$) & & & & & \\
\hline
\bottomrule
\end{tabular}
}
\end{table}

Overall, DIBmix demonstrated strong and consistent performance across diverse data sets, achieving the highest ARI values in four out of the ten cases. Importantly, a clear pattern emerges:

\begin{itemize}
\item Best performance occurs when data sets contain a balanced mix of variable types. For instance, in the Heart Disease, Adult, Statlog, and Australian Institute of Sport data sets, which all feature a substantial but not overwhelming proportion of categorical variables, DIBmix outperformed its competitors. 
\item Performance decreases at the extremes. When data sets are dominated by categorical variables (e.g., Dermatology, with nearly 97\% categorical features), or continuous (e.g., Australian Institute of Sport, Echocardiogram, and Byar Prostate Cancer with 92\%, 78\%, and 62\% continuous features, respectively), other methods such as FAMD or K-Prototypes performed equally well or better.
\end{itemize}

\begin{figure}[h!]
\centering
	\includegraphics[scale=0.5]{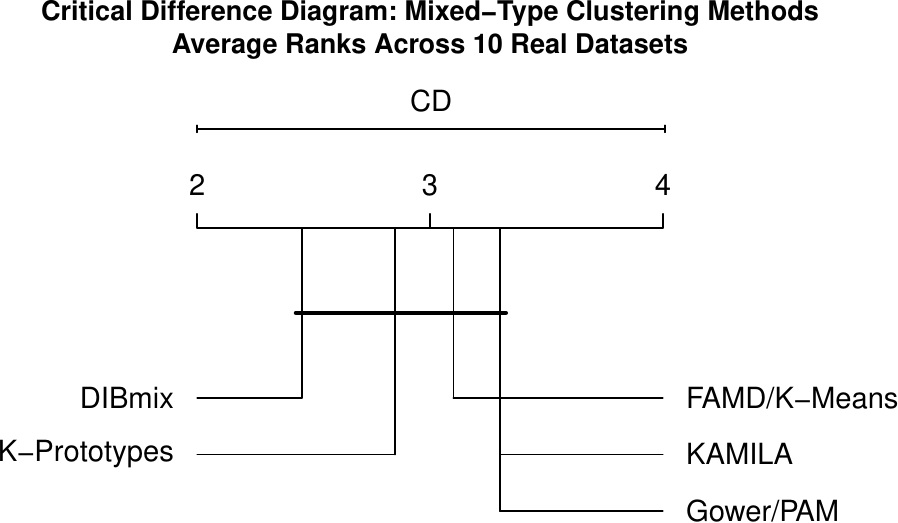}
	\caption{Critical Difference (CD) diagram for five clustering methods on 10 publicly available data sets. Average ranks are shown with lower values indicating better performance. The critical difference (CD = 1.929) indicates the minimum rank difference for significance at $\alpha = 0.05$.}
	\label{fig:CDplot_dibmix} 
\end{figure}

Even in cases where DIBmix was not the top method, it produced non-zero and often competitive ARI values, extracting meaningful cluster structure. For example, in the Inflammation and Credit Approval data sets, it outperformed some but not all competitors. \textcolor{black}{We performed a Friedman test on the results for the ten publicly available datasets at 95\% significance level. The null hypothesis of equal ranks is not rejected but the critical difference (CD = 1.929; see Figure \ref{fig:CDplot_dibmix}) indicates that methods must differ by approximately two full rank positions to achieve statistical significance. Even the largest observed difference (DIBmix at rank 2.45 vs. Gower/PAM at rank 3.30) represents less than half this threshold. With only ten datasets spanning diverse applications, sample sizes ($n=61$ to $30162$), dimensionalities ($p=6$ to $34$), cluster structures ($C=2$ to $6$), and proportions of categorical variables ($8.33$\% to $97.06$\%), achieving statistical significance would require one method to dominate nearly all comparisons. Despite this, DIBmix demonstrates practical advantages: best average rank (2.45), top performance on four datasets, and robust results with no near-zero ARI values (minimum ARI = 0.1739). This consistency establishes DIBmix as a competitive clustering method for mixed-type data.}

Taken together, these results suggest that DIBmix is particularly well-suited to real-world applications involving genuinely mixed data, where neither categorical nor continuous features dominate. At the same time, its relative weaknesses at the extremes indicate that complementary methods may remain preferable in data sets that are overwhelmingly categorical or continuous.

\textcolor{black}{It is important to stress that despite its poor performance on certain data sets in terms of ARI, the clustering solution need not be assessed solely on its classification performance. While it is common for classification data sets to be used for benchmarking clustering techniques, the performance of a clustering algorithm is mainly assessed by interpreting its clusters \cite{hennig2015true}. In fact, the partition that is defined by the labels of a classification data set is usually less optimal than the one obtained using a clustering algorithm, with respect to some desirable properties that a cluster has to possess \cite{bautista2025ground}.}

\begin{figure}[h!]
\centering
	\includegraphics[scale=0.6]{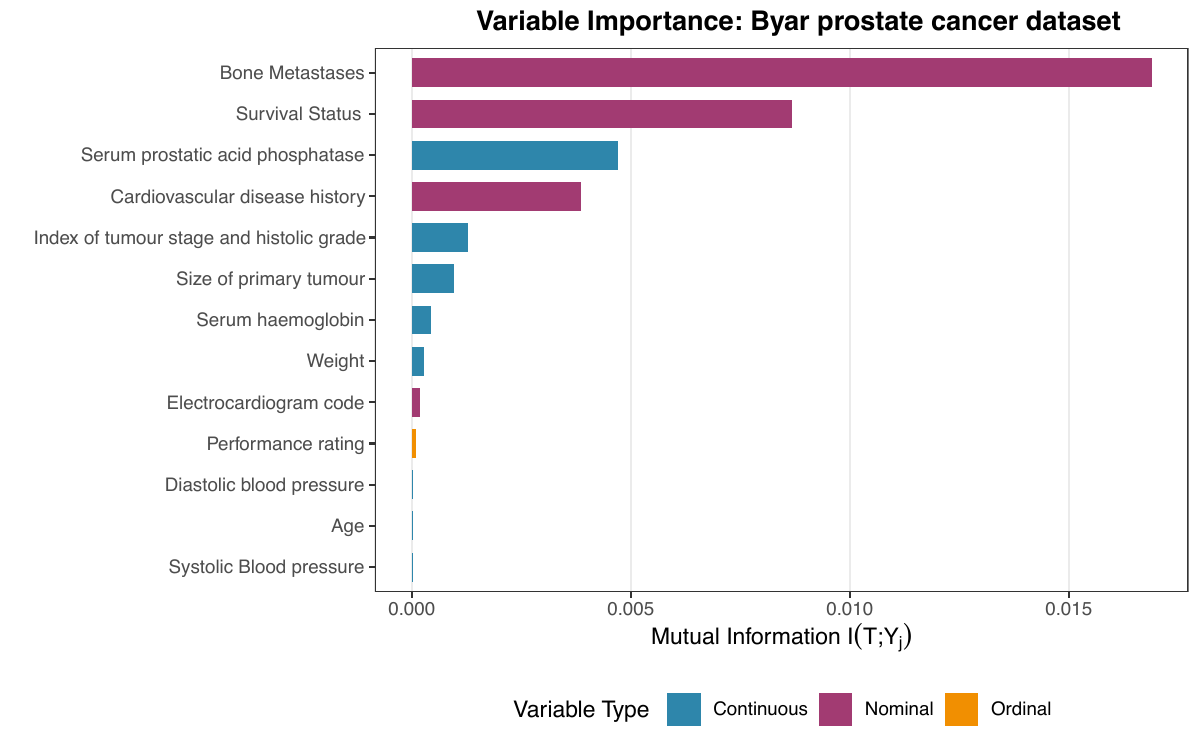}
	\caption{Post-analysis of the Byar prostate cancer data set. Variable importance is quantified by the mutual information between each feature and the partition obtained by DIBmix.}
	\label{fig:Byar_postanalysis} 
\end{figure}

\textcolor{black}{A desirable property of DIBmix is that it addresses this interpretive need through variable importance quantification. While bandwidth selection is done so as to ensure that no variable dominates the clustering solution a priori, the proposed algorithm learns a partition in which certain variables naturally emerge as more informative for distinguishing between clusters. Variable importance can be assessed by computing the mutual information between the clustering output and each individual feature, the distribution of the latter being estimated using the input kernel function and the bandwidth value that was determined by DIBmix. For instance, Figure \ref{fig:Byar_postanalysis} shows that clinically meaningful features like bone metastases, the post-trial survival status of patients, and a prostate-specific biomarker (serum prostatic acid phosphatase) drove the obtained partition, while age or blood pressure measurements contributed minimally.}

\textcolor{black}{The runtime analysis provided in \ref{appen6} indicates that DIBmix executes efficiently and is sometimes even quicker than its competitors on data sets with few observations. For large data sets, such as the Adult/Census Income, the Nystr\"{o}m approximation is essential; without it, DIBmix needed over eleven hours on this data set. Finally, while the hyperparameter search increases computational cost, the computational time remains within practical limits.}

\section{\textcolor{black}{Selecting the Number of Clusters}}\label{Sec:num_clusters}

\textcolor{black}{Estimating the number of clusters is a notoriously difficult problem in cluster analysis. Here, we present information-theoretic implementations of two common approaches: the elbow method and the Gap statistic \cite{tibshirani2001estimating}.}

\textcolor{black}{The mutual information \( I(T; Y) \) increases with the number of clusters (see Proposition \ref{prop:infoincrease} and Lemma \ref{lemma:usefulres} in \ref{appen2} and \ref{appen3}). Experiments on synthetic data show that the rate of increase is greatest when transitioning from \( C^*-1 \) to \( C^* \) clusters, where \( C^* \) is the true number of clusters. We tested this on 100 data sets with four well-separated spherical clusters and 100 with moderately overlapping non-spherical clusters, each containing 500 observations and eight variables (four continuous, four categorical). The knee of the mutual information curve correctly identified $C=4$ in 97\% of well-separated cases and 85\% of overlapping cases (Figure \ref{fig:kneeplots}, \ref{appen4}).}

\textcolor{black}{For the Gap statistic, we compute $\text{Gap}(C) = I(T_C; Y) - \sum_{b=1}^B I(T_C; Y_b)/B$ using $B=100$ reference data sets generated by independent variable permutation. Selecting $C$ based on maximum Gap yields the correct number of clusters in 80\% of well-separated cases but only 25\% with moderate overlap, as the Gap curve plateaus due to the monotonic increase of $I(T; Y)$. The `one standard error' rule improves this to 84\% and 34\%, respectively (Figure \ref{fig:Gapplots}, \ref{appen5}).}

\textcolor{black}{Overall, the knee heuristic outperforms the Gap statistic for DIBmix and offers a promising direction for cluster validation in mixed-type data, where few suitable methods currently exist \cite{arbelaitz2013extensive}.}

\section{Conclusion} \label{Sec:conclusion}

\textcolor{black}{In this paper, we introduced DIBmix, a clustering method for mixed-type data based on the Deterministic Information Bottleneck framework. We described its mathematical foundations, analysed the role of its hyperparameters, and proposed a strategy to select them so that different variable types contribute equally to the final partition. By recursively updating the regularisation parameter $\beta$, the method guarantees a valid partition into exactly $C$ clusters. Simulation studies and applications to publicly available datasets show that DIBmix performs competitively with state-of-the-art clustering methods for heterogeneous data. The framework is flexible and can be extended to purely continuous or categorical data, as well as to other data types such as counts through appropriate kernel smoothing techniques \cite{kokonendji2007discrete}.}

\textcolor{black}{Despite its strong performance, DIBmix has several limitations. Bandwidth selection is critical for detecting cluster structure, and while the proposed approach performs well overall, it may fail in certain scenarios. In addition, the method implicitly assumes cluster homogeneity, which can be problematic when clusters differ substantially in size or compactness, making the use of a single bandwidth suboptimal. Future work could address this issue by incorporating local or cluster-specific information, for example by borrowing ideas from nearest-neighbour methods \cite{hastie1995discriminant}.}

\textcolor{black}{Nevertheless, DIBmix provides a flexible framework for information-based clustering of heterogeneous data and opens several directions for future research. These include feature weighting and selection using the spectral properties of the perturbed matrix $\mathbf{P'}$, although the associated computational costs remain prohibitive in high-dimensional settings. Another promising direction concerns robustness; extending DIBmix to allow observation-specific weights, for instance through trimming strategies similar to Trimmed K-Means \cite{cuesta1997trimmed}, could improve performance in the presence of outliers.}

\newpage
\appendix
\section{The discrete uniform is a maximum entropy distribution}\label{appen1}
\begin{theo}\label{thm:max_entropy}
The discrete uniform distribution with support $\mathcal{S}$ is the maximum entropy distribution among all discrete random variables with the same support. 
\end{theo}
\begin{pf}

Let $X$ be a discrete uniform random variable with support $\mathcal{S} = \{ 1, \ldots, C\}$, where $C \in \mathbb{N}$. Then the probability mass function of $X$ is given by $\mathbb{P}(X=x) 
= 1/C \forall x \in \mathcal{S}$. Now let $Y$ be a discrete random variable defined on the same support $\mathcal{S}$. Define the distributions of $X$ and $Y$ by $Q$ and $P$, respectively. Then, the Kullback-Leibler divergence between $P$ and $Q$ in bits is given by:

\begin{align*}
    0 \leq D_\text{KL}(P \lvert\rvert Q) & = \sum_{x \in \mathcal{S}} P(X)\log_2{\left( \frac{P(x)}{Q(x)}\right)} \\
    & = \sum\limits_{x \in \mathcal{S}} P(X)\log_2{P(X)} - \sum\limits_{x \in \mathcal{S}} P(X)\log_2{Q(X)}\\
    & = -H(P) - \sum\limits_{x \in \mathcal{S}} P(X)\log_2{Q(X)},
\end{align*}
where $H(P)$ is the entropy of $P$ in bits. Then, we can further simplify the second term, since we know that $Q(X) = 1/C \forall x \in \mathcal{S}$, which gives:
\begin{align*}
    D_\text{KL}(P \lvert\rvert Q) & = -H(P) - \log_2{\left(\frac{1}{C}\right)}\sum\limits_{x\in \mathcal{S}} P(x)\\
    & = -H(P) - \log_2{\left(\frac{1}{C}\right)}.
\end{align*}
Finally, it suffices to see that the entropy of $Q$, that is the entropy of the discrete uniform distribution on $\mathcal{S}$, is in fact equal to the second term:
\begin{align*}
    H(Q) & = -\sum\limits_{x\in \mathcal{S}} Q(x)\log_2{Q(x)}\\
    & = -\log_2{\left(\frac{1}{C}\right)} \sum\limits_{x\in \mathcal{S}} Q(x)\\
    & = -\log_2{\left(\frac{1}{C}\right)}.
\end{align*}
Combining everything, we get:
\begin{align*}
    0 \leq D_\text{KL}(P \lvert\rvert Q) & = -H(P) - \log_2{\left(\frac{1}{C}\right)}\\
    & = - H(P) + H(Q),
\end{align*}
which gives $H(Q) \geq H(P)$, proving that the discrete uniform distribution is a maximum entropy distribution with support $\mathcal{S}$.
\end{pf}

\section{Mutual Information increases when a cluster is split}\label{appen2}
\begin{prop}\label{prop:infoincrease}
    Let $\mathcal{T}$ and $\mathcal{T}'$ be two partitions into $C$ and $C+1$ clusters respectively obtained by DIBmix. Assuming that the additional cluster in $\mathcal{T}'$ is obtained by splitting one of the $C$ clusters of $\mathcal{T}$, it holds that $I(Y; \mathcal{T}) < I(Y; \mathcal{T}')$.
\end{prop}

\begin{pf}
    We begin by considering the definition of mutual information. More precisely, for any partition $T$:
    $$I(Y; T) = H(Y) - H(Y \mid T),$$
    where $H(Y \mid T)$ is the conditional entropy of the location of observations $Y$ given their cluster assignment $T$. Notice that the conditional entropy can be rewritten as the following weighted sum:
    $$H(Y \mid T) = \sum_{t \in T} \mathbb{P}(T=t)H(Y \mid T = t).$$
    Given that $H(Y)$ is independent of the partition obtained and assuming without loss of generality that the $C$th and the $(C+1)$st clusters in $\mathcal{T}'$ emerge by splitting the $C$th cluster of $\mathcal{T}$, our problem is equivalent to that of proving the following:
    $$\mathbb{P}(\mathcal{T}' = C)H(Y \mid \mathcal{T}' = C) + \mathbb{P}(\mathcal{T}' = C+1)H(Y \mid \mathcal{T}' = C+1) < \mathbb{P}(\mathcal{T} = C)H(Y \mid \mathcal{T} = C).$$
    Notice that $\mathbb{P}(\mathcal{T} = C) = \mathbb{P}(\mathcal{T}' = C) + \mathbb{P}(\mathcal{T}' = C+1)$, hence it suffices to show that $H(Y \mid \mathcal{T}' = C) < H(Y \mid \mathcal{T} = C)$ and $H(Y \mid \mathcal{T}' = C+1) < H(Y \mid \mathcal{T} = C)$. Let us begin by considering $H(Y \mid \mathcal{T}' = C+1)$. For the sake of simplicity, assume without loss of generality that the first $n_C$ observations were assigned into the $C$th cluster in $\mathcal{T}$ and the first $n_{C+1}' < n_C$ of them are then assigned in the $(C+1)$st cluster in $\mathcal{T}'$. What this means is that for observations $x$ for which $q(\mathcal{T}' = C+1 \mid x) = 1$ and ensuring that the value of $\beta$ is large enough to ignore the contribution of the compression/entropy term, the following inequality needs to hold:
    \begin{equation}\label{eq:dkl_newclust}
    D_{\text{KL}}\left( p(y \mid x) \lvert\rvert q(y \mid \mathcal{T}' = C+1) \right) < D_{\text{KL}}\left( p(y \mid x) \lvert\rvert q(y \mid \mathcal{T}' = t) \right) \forall t \in \{1, \ldots, C\}.
    \end{equation}
    We know that the clustering is being done based on the perturbed similarity matrix $\mathbf{P'}$, with observations for which the corresponding $p(y \mid x)$ values are large being more likely to be assigned in the same cluster. Hence, the entries of the block $\mathbf{P'}_{[(1:n'_{C+1}) \times (1:n'_{C+1})]}$ will take larger values than these of $\mathbf{P'}_{[(1:n'_{C+1}) \times (n'_{C+1}:n_C)]}$. Consequently, for Expression \eqref{eq:dkl_newclust} to hold and since the first $n'_{C+1}$ elements are the ones for which  $q(\mathcal{T}' = C+1 \mid x) = 1$, the first $n'_{C+1}$ elements of $q(y \mid \mathcal{T}' = C+1)$ must be larger than the first $n'_{C+1}$ elements of $q(y \mid \mathcal{T}' = C)$. However, if the first $n'_{C+1}$ elements of $q(y \mid \mathcal{T}' = C+1)$ increase, then the remaining $n - n'_{C+1}$ elements must decrease, so as to ensure that all elements of $q(y \mid \mathcal{T}' = C+1)$ sum to a unit. This leads to the distribution of $q(y \mid \mathcal{T}' = C+1)$ being more skewed than that of $q(y \mid \mathcal{T} = C)$ which implies that $H(Y \mid \mathcal{T}' = C+1) < H(Y \mid \mathcal{T} = C)$. Similarly, it can be shown that $H(Y \mid \mathcal{T}' = C) < H(Y \mid \mathcal{T} = C)$. Combining everything, we obtain the desired result. This also gives us some interesting mathematical properties of the perturbed similarity matrix $\mathbf{P}'$ which are outlined in Lemma \ref{lemma:usefulres}.
\end{pf}

\section{Perturbed similarity matrix entry properties}\label{appen3}
\begin{lemma}\label{lemma:usefulres}
    Let $\mathcal{T}$ and $\mathcal{T}'$ be two partitions into $C$ and $C+1$ clusters respectively obtained by DIBmix. Assume without loss of generality that the additional cluster is obtained by splitting the $C$th cluster of $\mathcal{T}$ which includes $n_C$ observations and that the two resulting clusters include $n_C'$ and $n_{C+1}'$ observations each. Then, for any observation $i \in \{x: q(\mathcal{T}' = C \mid x) = 1 \}$, it holds that:
    $$\frac{\sum\limits_{j : q(\mathcal{T}' = C \mid j) = 1} \mathbf{P'}_{ij}}{\sum\limits_{j : q(\mathcal{T}' = C+1 \mid j) = 1} \mathbf{P'}_{ij}} > \frac{n_C'}{n'_{C+1}}$$
    and for any observation $i \in \{x: q(\mathcal{T}' = C \mid x) = 0 \}$:
    $$\frac{\sum\limits_{j : q(\mathcal{T}' = C \mid j) = 1} \mathbf{P'}_{ij}}{\sum\limits_{j : q(\mathcal{T}' = C+1 \mid j) = 1} \mathbf{P'}_{ij}} < \frac{n_C'}{n'_{C+1}}.$$
\end{lemma}

\begin{pf}
    This result can easily be derived by looking at Proposition \ref{prop:infoincrease}. More precisely, we have that:
    $$q(y \mid T = t) = \frac{1}{\sum\limits_{i} q(T = t\mid i)} \times \sum\limits_{i} q(T = t\mid i)\mathbf{P'}_{ij}.$$
    Now for $i \in \{x: q(\mathcal{T}' = C \mid x) = 1 \}$, based on Proposition \ref{prop:infoincrease}:
    $$[q(y \mid \mathcal{T}' = C)]_i > [q(y \mid \mathcal{T} = C)]_i.$$
    Therefore, by considering the definition of $q(y \mid T = t)$, the above Expression becomes:
    $$\frac{1}{n_C'}\sum\limits_{j : q(\mathcal{T}' = C \mid j) = 1} \mathbf{P'}_{ij} > \frac{1}{n_C}\sum\limits_{j : q(\mathcal{T} = C \mid j) = 1} \mathbf{P'}_{ij}.$$
    Notice that the sum on the right hand side can be decomposed as follows:
    $$\sum\limits_{j : q(\mathcal{T} = C \mid j) = 1} \mathbf{P'}_{ij} = \sum\limits_{j : q(\mathcal{T}' = C \mid j) = 1} \mathbf{P'}_{ij} + \sum\limits_{j : q(\mathcal{T}' = C+1 \mid j) = 1} \mathbf{P'}_{ij}.$$
    Therefore, we get:
    \begin{align*}
        \frac{1}{n_C'}\sum\limits_{j : q(\mathcal{T}' = C \mid j) = 1} \mathbf{P'}_{ij} & > \frac{1}{n_C}\sum\limits_{j : q(\mathcal{T}' = C \mid j) = 1} \mathbf{P'}_{ij} + \frac{1}{n_C} \sum\limits_{j : q(\mathcal{T}' = C+1 \mid j) = 1} \mathbf{P'}_{ij}\\
        \iff \frac{n_C-n'_C}{n_cn_C'}\sum\limits_{j : q(\mathcal{T}' = C \mid j) = 1} \mathbf{P'}_{ij} & > \frac{n'_C}{n_Cn'_C} \sum\limits_{j : q(\mathcal{T}' = C+1 \mid j) = 1} \mathbf{P'}_{ij}\\
        \iff \frac{\sum\limits_{j : q(\mathcal{T}' = C \mid j) = 1} \mathbf{P'}_{ij}}{\sum\limits_{j : q(\mathcal{T}' = C+1 \mid j) = 1} \mathbf{P'}_{ij}} &> \frac{n'_C}{n_C-n'_C}.
    \end{align*}
    Finally, notice that the $C$th and the $(C+1)$st clusters from $\mathcal{T}'$ partition the $C$th cluster from $\mathcal{T}$, hence $n'_C + n'_{C+1} = n_C$, leading us to the required expression. The analogous result for observations $i \in \{x: q(\mathcal{T}' = C \mid x) = 0 \}$ can be easily shown in a similar manner, by considering that:
    $$[q(y \mid \mathcal{T}' = C)]_i < [q(y \mid \mathcal{T} = C)]_i.$$
\end{pf}

\section{Results of the knee heuristic simulations}\label{appen4}

\begin{figure}[H]
\centering
	\includegraphics[scale=0.8]{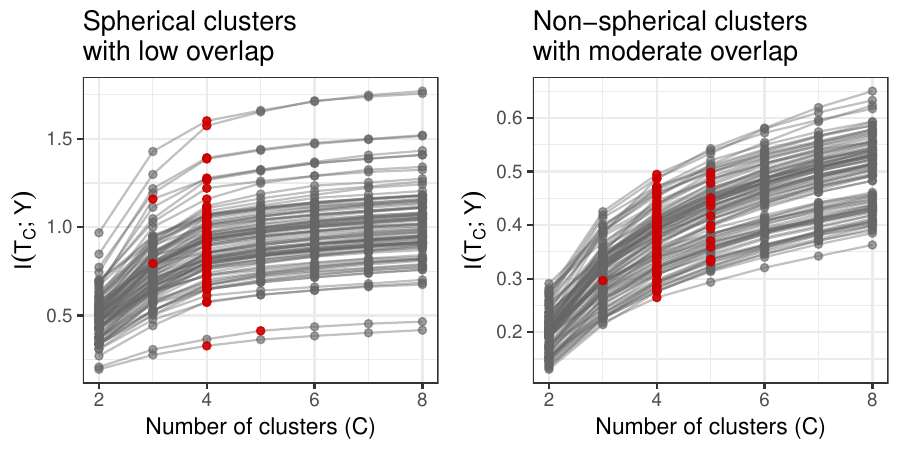}
	\caption{Mutual information curves against the number of clusters $C$ that DIBmix is run with for synthetic data sets with four well-separated spherical and moderately-separated non-spherical clusters, respectively. The red points correspond to the knee points of each of the curves.}
	\label{fig:kneeplots} 
\end{figure}

\section{Results of the Gap statistic simulations}\label{appen5}

\begin{figure}[H]
\centering
	\includegraphics[scale=0.5]{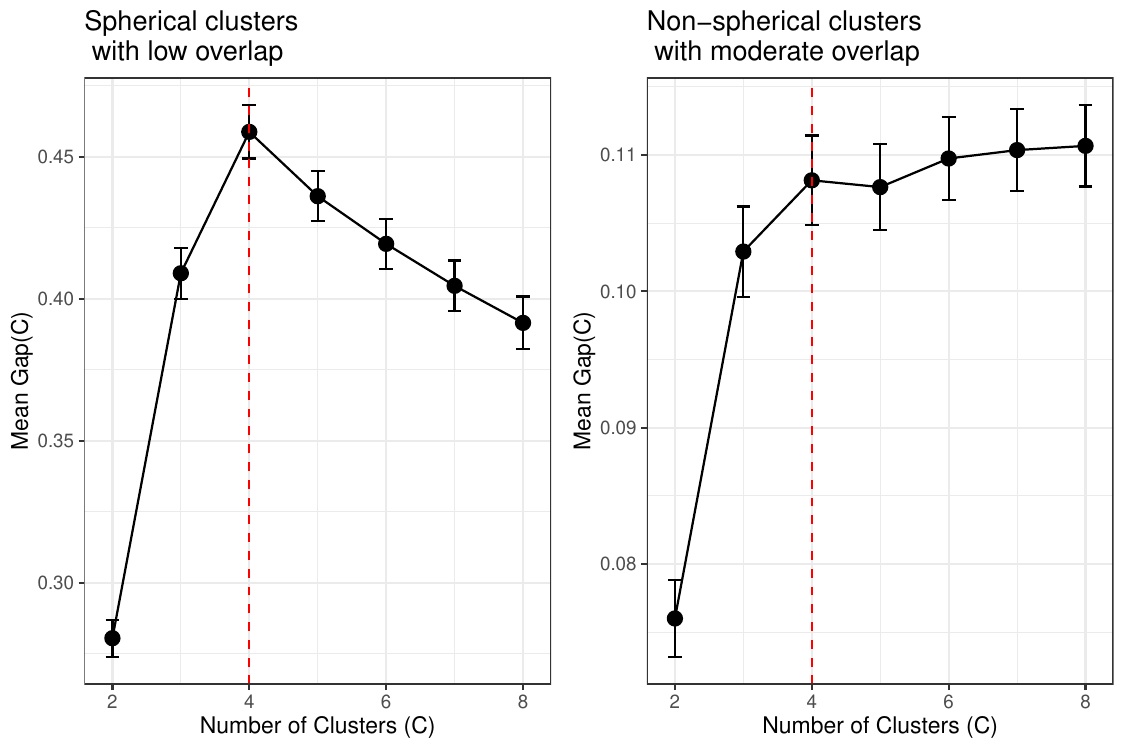}
	\caption{Average Gap statistic values across a hundred replicates for spherical clusters with low overlap and non-spherical clusters with moderate overlap. The dashed line indicates the true number of clusters and error bars represent the average standard deviation of the Gap statistic for each number of clusters $C$.}
	\label{fig:Gapplots} 
\end{figure}
\vspace{-0.2cm}
\section{Runtime analysis}\label{appen6}
\vspace{-0.2cm}
\begin{table}[!ht]
\caption{Runtimes (in seconds) of five clustering methods on ten mixed-type data sets from the UCI repository. Greatest runtimes for each data set appear in bold.}
\label{tab:runtimes}
\vspace{-0.2cm}
\scriptsize{
\begin{tabular}{lrrrrrr}
\toprule
\hline
Dataset & DIBmix$_{\text{search}}$ & DIBmix$_{\text{fixed}}$ & KAMILA & K-Prototypes & FAMD & Gower/PAM \\
\hline
\midrule
Dermatology & \textbf{41.73} & 37.52 & 2.30 & 36.36 & 0.06 & 0.03\\
Heart Disease & \textbf{9.05} & 6.51 & 0.79 & 6.95 & 0.02 & 0.02\\
\textcolor{black}{Adult/Census Income} & \textcolor{black}{263.13} & \textcolor{black}{23.27} & 15.82 & \textbf{584.27} & 1.67 & 139.52 \\
Hepatitis & 1.57 & 0.59 & 0.64 & \textbf{5.80} & 0.02 & 0.01 \\
Australian Institute of Sport & 3.74 & 2.53 & 0.63 & \textbf{5.26} & 0.01 & 0.02\\
Inflammation & 1.52 & 0.58 & 0.42 & \textbf{2.31} & 0.08 & 0.01\\
Statlog (Australian) & \textbf{61.00} & 48.54 & 0.91 & 12.54 & 0.03 & 0.08\\
Credit Approval & \textbf{36.32} & 25.38 & 0.83 & 12.64 & 0.03 & 0.08\\
Echocardiogram & 1.31 & 0.39 & 0.59 & \textbf{3.46} & 0.01 & 0.01\\
Byar Prostate Cancer & \textbf{21.46} & 17.46 & 0.82 & 13.75 & 0.02 & 0.04\\
\hline
\bottomrule
\end{tabular}
}
\end{table}
\vspace{-0.2cm}
The runtimes reported in Table \ref{tab:runtimes} were recorded by running simulations on Imperial College London's CX3 HPC facility with AMD EPYC 7742 processors (2.25 GHz, 64 cores per processor). `DIBmix$_{\text{search}}$' and `DIBmix$_{\text{fixed}}$' correspond to DIBmix with bandwidths selected using the approach from Section \ref{sec:bwparams} and with fixed user-input bandwidths, respectively. The Nystr\"{o}m approximation with $m = \lceil \sqrt{30162} \rceil = 174$ landmark points and the data subsampling procedure for bandwidth selection (a hundred subsamples of a thousand observations each) were used on Adult/Census Income.
\vspace{-0.2cm}
\section{Effect of number of landmark points on Nystr\"{o}m approximation}\label{appen7}
\vspace{-0.2cm}
\begin{figure}[H]
\centering
	\includegraphics[scale=0.5]{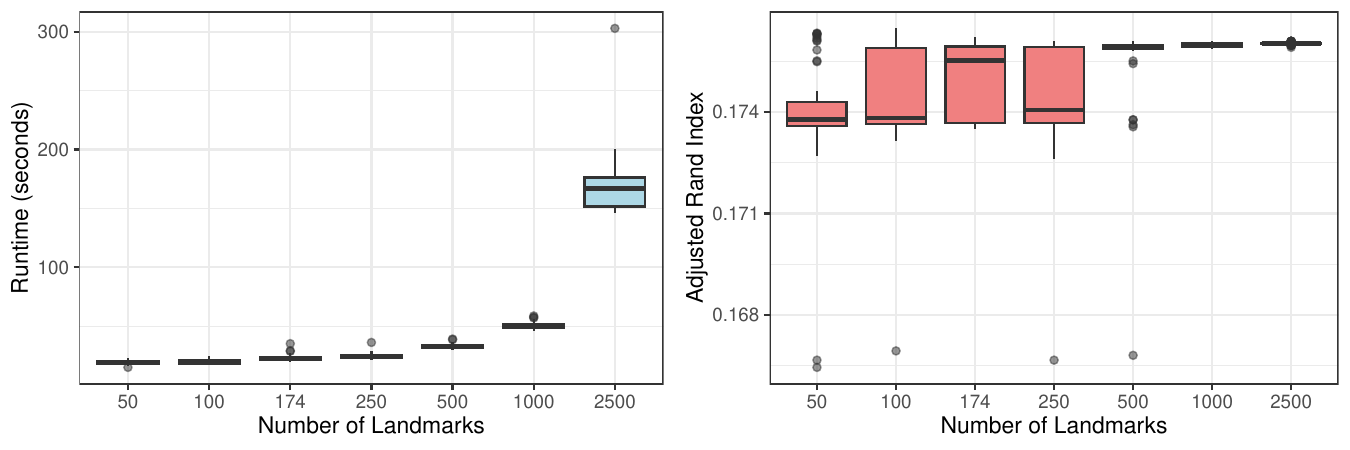}
    \vspace{-0.2cm}
	\caption{Distribution of runtimes (in seconds) and cluster recovery performance (in ARI) of DIBmix without hyperparameter selection on the Adult/Census Income data set ($n = 30162$) for $m = 50, 100, 174, 250, 500, 1000,$ and $2500$ random landmark points used in Nystr\"{o}m approximation. DIBmix was run fifty times with a hundred random initialisations and $m = \lceil \sqrt {n} \rceil = 174$ was eventually selected.}
	\label{fig:nystrom_runtime} 
\end{figure}

\section*{Funding}
The first author gratefully acknowledges funding provided by EPSRC's StatML CDT grant EP/S023151/1.

\section*{Conflict of interest}
The authors declare that they have no conflict of interest.

\section*{Data availability}
The extended results, the data sets obtained from the UCI repository, and the code to reproduce the analyses are all available online at \url{https://github.com/EfthymiosCosta/IBclust_Simulations}.

\bibliographystyle{elsarticle-num-names}
\bibliography{references.bib}

\end{document}